\documentclass[%
 reprint,floatfix,
superscriptaddress,
 amsmath,amssymb,
 aps,
]{revtex4-1}

\usepackage{graphicx}
\usepackage{dcolumn}
\usepackage{bm}
\usepackage{braket}
\usepackage{color}
\usepackage{latexsym}
\usepackage{amsmath,amsthm}
\usepackage{amsfonts}
\usepackage{amsmath,amssymb,amsthm,mathrsfs,amsfonts,dsfont}
\usepackage{subfigure, epsfig}
\usepackage{braket}
\usepackage{ulem}
\usepackage{bm}
\usepackage{enumerate}
\usepackage{color}
\usepackage{graphicx}

\begin{document}


\title{
Generation of multipartite entanglement
between spin-1 particles with bifurcation-based quantum annealing
}



 \author{Yuichiro Matsuzaki}
 \email{matsuzaki.yuichiro@aist.go.jp}
\affiliation{%
 Research Center for Emerging Computing Technologies, National Institute of Advanced Industrial Science and Technology (AIST),
1-1-1 Umezono, Tsukuba, Ibaraki 305-8568, Japan.
}%
\affiliation{
NEC-AIST Quantum Technology Cooperative Research Laboratory,
National Institute of Advanced Industrial Science and Technology (AIST), Tsukuba, Ibaraki 305-8568, Japan
}

\author{Takashi Imoto}
\affiliation{%
 Research Center for Emerging Computing Technologies, National Institute of Advanced Industrial Science and Technology (AIST),
1-1-1 Umezono, Tsukuba, Ibaraki 305-8568, Japan.
}%
\author{Yuki Susa}%
\affiliation{
NEC-AIST Quantum Technology Cooperative Research Laboratory,
National Institute of Advanced Industrial Science and Technology (AIST), Tsukuba, Ibaraki 305-8568, Japan
}
\affiliation{System Platform Research Laboratories, NEC Corporation, Kawasaki, Kanagawa 211-8666, Japan}

\date{\today}

\begin{abstract}
Quantum annealing is a way to solve a combinational optimization problem where quantum fluctuation is induced by transverse fields. Recently, a bifurcation-based quantum annealing with spin-1 particles
was suggested as another mechanism to implement the quantum annealing. 
In the bifurcation-based quantum annealing,
each spin is initially prepared in $|0\rangle$, let this state evolve by a time-dependent Hamiltonian in an adiabatic way, and we find a state spanned by $|\pm 1\rangle$
 at the end of the evolution.
Here, we propose a scheme to generate multipartite entanglement, namely GHZ states, 
between spin-1 particles
by using the bifurcation-based quantum annealing. We gradually
decrease the detuning of the spin-1 particles while we adiabatically change the amplitude of the external driving fields.
Due to the dipole-dipole interactions between the spin-1 particles, we can prepare the GHZ state after performing this protocol. We discuss possible implementations of our scheme by using
nitrogen vacancy centers in diamond.
\end{abstract}

\maketitle

\section{Introduction}

Quantum annealing (QA) is a technique for solving combinational optimization problems \cite{kadowaki1998quantum, farhi2000quantum, farhi2001quantum}.
The solution of the combinational optimization problems is embedded in the ground state of the Ising Hamiltonian \cite{lucas2014ising}, which is called the problem (or target) Hamiltonian. We use the transverse magnetic fields to induce quantum fluctuation, and this Hamiltonian is called the driving Hamiltonian.
After preparing a ground state of the driving Hamiltonian, 
we gradually decrease the amplitude of the transverse driving fields while we slowly increase the strength of the Ising Hamiltonian. If the dynamics is adiabatic, the ground state of the problem Hamiltonian can be prepared \cite{morita2008mathematical}.
Previous studies mainly focus on the use of two-level systems  for QA \cite{santoro2002theory,johnson2011quantum,boixo2013experimental,boixo2014evidence}.

The other mechanisms using bifurcation were proposed to induce the quantum fluctuations for QA. 
It is known that a parametrically driven Kerr nonlinear oscillator (KPO) shows the bifurcation \cite{wielinga1993quantum}. A quantum superposition of two distinct states of the KPO can be generated by using quantum adiabatic evolution through its bifurcation point.
Moreover, we can use this system as a qubit for a gate type-quantum computer \cite{cochrane1999macroscopically}.
Previous researches reveal that we can use the KPO for QA to find a ground state of Ising Hamiltonians \cite{goto2016bifurcation,puri2017quantum}.

Recently, Takahashi shows that we can use spin-1 particles for the bifurcation-based QA \cite{takahashi2020bifurcation}. 
For non-interacting spin-1 systems, the initial state is $|0\rangle $, and degenerate states $|\pm 1\rangle $ are prepared 
at the end of the evolution, which is similar to the bifurcation mechanism of the KPO.
On the other hand, 
for interacting spin-1 systems, the problem Hamiltonian is encoded in a subspace spanned by $|\pm 1\rangle $.
Each spin-1 particle is initially prepared in  $|0\rangle $, and adiabatic changes of the Hamiltonian including the coupling between the spin-1 particles
provide a ground state of the problem Hamiltonian \cite{takahashi2020bifurcation}.


Here, we propose a scheme to generate the GHZ states between spin-1 particles by using the bifurcation-based QA. 
Suppose that there are dipole-dipole interactions between the spin-1 particles. By choosing suitable parameters, the GHZ states have the lowest energy.
This means that, starting from a trivial ground state of $|00\cdots 0\rangle$ with longitudinal fields, we adiabatically change the Hamiltonian, and we can obtain the GHZ states where we add external transversal fields in the middle of the dynamics. 
Importantly, due to the degeneracy of the ground states of the target Hamiltonian, the energy gap between the ground state and excited states becomes small during QA.
However, we show that the total Hamiltonian commutes with a parity operator, and this symmetry can suppress the non-adiabatic transitions during QA.
Although this kind of the symmetry protected mechanism was discussed in the conventional QA \cite{xing2016heisenberg,hatomura2019superadiabatic,hatomura2019suppressing,huang2018non,hatomura2021symmetry}, we firstly utilize the symmetry protected mechanism
for the bifurcation-based QA. \textcolor{black}{Moreover, as a possible implementation, we discuss the use of nitrogen vacancy (NV) centers in diamond, and they are spin-1 particles that are candidates to realize quantum information processing.}

The paper is structured as follow. In section II, we review the conventional QA and bifurcation-based QA 
to find a ground state of the Ising Hamiltonian. In section III, we review the NV ceners in diamond. In section IV, we introduce our scheme to generate the GHZ states with the bifurcation-based QA. In section V, we perform numerical simulations to evaluate the performance of our scheme. In section VI, we summarize our results.

\section{Quantum annealing}
\subsection{Conventional quantum annealing with spin-1/2 particles}
Here, we review the conventional QA with spin-1/2 particles \cite{kadowaki1998quantum, farhi2000quantum, farhi2001quantum}.
The main aim of QA is to prepare a ground state of the following Ising-type Hamiltonian.
\begin{eqnarray}
 H^{(1/2)}_{\rm{P}}=\sum _{j=1}^L h_j \hat{\sigma}_z^{(j)}+ \sum _{ i\neq j} J_{i,j}\hat{\sigma}_z^{(i)}\hat{\sigma}_z^{(j)}
\end{eqnarray}
where $L$ denotes the number of spins, $h_j$ denotes a longitudinal field at the $j$-th spin, and $J_{i,j}$ denotes the coupling strength between the \textcolor{black}{$i$-th spin and $j$-th spin}.
We also use a driver Hamiltonian to induce the quantum fluctuation as follows.
\begin{eqnarray}
 H^{(1/2)}_{\rm{D}}=\sum _{j=1}^L B_j \hat{\sigma}_x^{(j)}
\end{eqnarray}
where $B_j$ denotes transverse fields.
The total Hamiltonian is described as follows.
\begin{eqnarray}
 H^{(1/2)}=(1-t/T)H^{(1/2)}_{\rm{D}} + (t/T)H^{(1/2)}_{\rm{P}}
\end{eqnarray}
where $T$ denotes the time to implement QA.
In QA, we prepare a ground state of $H_{\rm{D}}$, and let this state evolve by the total Hamiltonian. It is known that,
as long as an adiabatic condition is satisfied, we can obtain a ground state of the total Hamiltonian.

\subsection{Bifurcation-based quantum annealing with spin-1 particles}
Let us review a bifurcation-based quantum annealing with spin-1 particles \cite{takahashi2020bifurcation}.
We consider the following driving Hamiltonian
\begin{eqnarray}
 H_{\rm{D}}=\sum_{j=1}^L A(t) \hat{S}_x^{(j)}+ C(t) (\hat{S}_z^{(j)})^2
\end{eqnarray}
where
$\hat{S}_x= |B\rangle \langle 0| +|0\rangle \langle B|$, 
$\hat{S}_y= -i|D\rangle \langle 0| +i|0\rangle \langle D|$, $\hat{S}_z=|1\rangle \langle 1|- |-1\rangle \langle -1|$, and $|B\rangle =\frac{1}{\sqrt{2}}(|+1\rangle +|-1\rangle )
$,
$|D\rangle =\frac{1}{\sqrt{2}}(|+1\rangle -|-1\rangle )
$.
We slowly change $C(t)$ from a positive large value to a negative large value while $A(t)$ has a finite but a small value in the middle of QA.
The problem Hamiltonian is given as
\begin{eqnarray}
 H_{\rm{P}}=\sum _{j=1}^L h_j \hat{S}_z^{(j)}+ \sum _{i\neq j}J_{i,j}\hat{S}_z^{(i)}\hat{S}_z^{(j)}
\end{eqnarray}
and the total Hamiltonian is given as
\begin{eqnarray}
 H=H_{\rm{D}}+H_{\rm{P}}.
\end{eqnarray}
We set $|C(0)|=|C(T)|\gg |h_j|, |J_{i,j}|$, and the ground state of the total Hamiltonian at $t=0$ is $\bigotimes _{j=1}^L|0\rangle _j$. By letting this state evolve by the total Hamiltonian, we obtain the ground state of the problem Hamiltonian as long as the adiabatic condition is satisfied.

\section{The nitrogen vacancy centers in diamond}
We review the Hamiltonian of the NV centers in diamond. The NV center is a spin-1 patricle, and there is a dipole-dipole interaction between the NV centers.
The Hamiltonian is described as follows
\begin{eqnarray}
&&H^{(\rm{NV})}=\sum _{j=1}^L\Big{(} D^{(j)}_0(\hat{S}_z^{(j)})^2
+ E_x^{(j)}(
(\hat{S}_x^{(j)})^2 - (\hat{S}_y^{(j)})^2
)
\Big{)}\nonumber \\
&+&\Big{(}
\sum_{j\neq k}J_{j,k}(\hat{S}_x^{(j)}\hat{S}_x^{(k)}
+\hat{S}_y^{(j)}\hat{S}_y^{(k)}
)- J' _{j,k}\hat{S}_z^{(j)} \hat{S}_z^{(k)}
\Big{)}\ \ \ .
\end{eqnarray}
where $D^{(j)}_0$ denotes a zero-field splitting at the $j$-th spin, $E_x^{(j)}$ denotes a strain at the $j$-th spin, $J_{j,k}$ denotes the flip-flop interaction between the $j$-th spin and  $k$-th spin, and $J'_{j,k}$ denotes the Ising interaction between the $j$-th spin and  $k$-th spin. It is worth mentining that we can change the values of $D^{(j)}_0$ ($E_x^{(j)}$)
by changing the temperature (amplitude of the applying electric fields) \cite{neumann2013high,clevenson2015broadband,dolde2011electric,kobayashi2020electrical,iwasaki2017direct}. 

The NV center is a promising candidate to realize quantum information processing.
The NV center can be coupled with magnetic fields, electric fields, and temperature, and pressure \cite{degen2017quantum,barry2020sensitivity,dolde2011electric}. 
We can polarize the NV centers by illuminating a green laser, and also we can readout the spin state by using the photoluminescence from the NV centers \cite{gruber1997scanning,degen2017quantum,barry2020sensitivity}.  Moreover, the NV center has a long coherence time such as a few milliseconds
 \cite{balasubramanian2009ultralong,mizuochi2009coherence,herbschleb2019ultra}.
 The NV center can be coherently coupled with an optical photon \cite{bernien2013heralded}.
\textcolor{black}{
These properties are prerequisite for the NV centers to be candidates for the quantum sensing \cite{degen2017quantum,maze2008nanoscale,taylor2008high,balasubramanian2008nanoscale}, a quantum memory for a superconducting qubit \cite{kubo2011hybrid,zhu2011coherent,saito2013towards,zhu2014observation}, quantum communication \cite{childress2005fault}, and distributed quantum computation \cite{nemoto2014photonic}.
Especially, NV centers could be used to realize an entanglement-enhanced quantum sensing with the GHZ states
\cite{wineland1992spin,wineland1994squeezed,huelga1997improvement,shaji2007qubit,matsuzaki2011magnetic,chin2012quantum,chaves2013noisy,isogawa2021vector} or could be used for a quantum network with encoding where the GHZ states are resource to construct an error correcting code
\cite{jiang2009quantum,satoh2012quantum,zwerger2016measurement}.}


\begin{figure}[h]
  \includegraphics[width=0.45\textwidth]{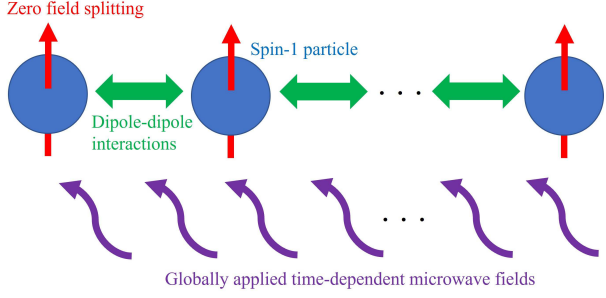}
  \caption{Schematic of our proposal. Spin-1 particles with zero-field splitting
  are arranged in a one-dimensional chain. There are dipole-dipole interactions between the spin-1 particles. We globally apply time-independent microwave fields to implement the bifurcation-based quantum annealing.
    }
  \label{schematicdiagram}
\end{figure}
\section{Generation of the GHZ states with bifurcation-based quantum annealing}
We explain our scheme to generate a GHZ state between spin-1 particles with the bifurcation-based QA. 
The schemetic is shown in Fig. \ref{schematicdiagram}.
We consider to apply our scheme with the NV centers in diamond. 
Importantly, in an experiment, it is difficult to have a negative value of $D^{(j)}_0$.
Although we can slightly change the value of $D^{(j)}_0$ by changing the temperature, the value of $D^{(j)}_0$ is as large as $2\pi \times 2.88$ GHZ, and there is no experiment to change the value of the zero field splitting to the negative values, which requires a frequency shift of a few GHZ. To overcome this problem, we adopt an idea of a spin-lock QA where the system driven by microwave fields is in a rotating frame \cite{chen2011experimental,matsuzaki2020quantum,imoto2021preparing}. The advantage of this scheme is that the detuning between the resonant frequency of the spins and the microwave frequency plays an role of the longitudinal fields, and we can easily set the negative detuning by setting a suitable value of the microwave frequency. 
\begin{figure}[h]
  \includegraphics[width=0.45\textwidth]{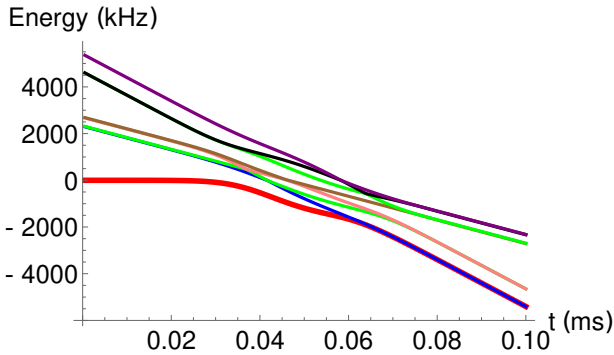}
  \caption{We plot eigenenergies
  against the time with the unit of milliseconds where we adopt the Hamiltonian in the Eq.\eqref{rwah}, which is
  derived by using a rotating wave approximation. 
   The energy gap between the ground state and first excited state becomes smaller as $t$ approaches to $T$.
  We set the parameters as $L=2$,
  $D_0'/2\pi= 400$ kHz, $E_x^{(1)}/2\pi= 1$ kHz, $E_x^{(1)}/2\pi= 1.2$ kHz, $J_{12}/2\pi=30$ kHz, $J_{12}'/2\pi=60$ kHz, $B/2\pi=100$ kHz, $\sigma =0.2 T$, $T=0.1$ ms.
    }
  \label{energydiagram}
\end{figure}
When the NV centers are arranged in a one dimensional chain and
microwave driving field
are applied along $x$ direction,
the Hamiltonian is described as follows.
\begin{eqnarray}
&&H=\sum _{j=1}^L\Big{(} D^{(j)}_0(\hat{S}_z^{(j)})^2
+2  \lambda ^{(j)}_x (t) \cos \omega t\  \hat{S}_x ^{(j)}
\nonumber \\
&+& E_x^{(j)}(
(\hat{S}_x^{(j)})^2 - (\hat{S}_y^{(j)})^2
)
\Big{)}\nonumber \\
&+&\Big{(}
\sum_{j\neq k}J_{j,k}(\hat{S}_x^{(j)}\hat{S}_x^{(k)}
+\hat{S}_y^{(j)}\hat{S}_y^{(k)}
)- J' _{j,k}\hat{S}_z^{(j)} \hat{S}_z^{(k)}
\Big{)}\  \label{hnorwa}
\end{eqnarray}
where $\lambda ^{(j)}_x (t)$ ($\omega$) denotes the amplitude (frequency) of the microwave driving at the $j$-th NV center. 
The dipole-dipole interactions decrease by $1/r^3$ where $r=|j-k|$ denotes the distance between the spins. For example,  $J_{13}=\frac{1}{8}J_{12}$ is satisfied.
In a rotating frame defined by
$U=e^{-i  \sum_{j=1}^{L}\omega (\hat{S}_z^{(j)})
}$, we obtain 
\begin{eqnarray}
&&H\simeq \sum _{j=1}^L\Big{(} D'_j(\hat{S}_z^{(j)})^2
+ \lambda ^{(j)}_x (t) \hat{S}_x ^{(j)}
\nonumber \\
&+& E_x^{(j)}(
(\hat{S}_x^{(j)})^2 - (\hat{S}_y^{(j)})^2
)
\Big{)}
+\Big{(}
\sum_{j\neq k}J_{j,k}(|B\rangle _j\langle 0|\otimes |0\rangle _{k}\langle B|\nonumber \\
&+&|D\rangle _j\langle 0|\otimes |0\rangle _{k}\langle D| + {\rm{hc}}
)
- J' _j\hat{S}_z^{(j)} \hat{S}_z^{(k)}\Big{)}
\label{rwah}
\end{eqnarray}
where we define 
$D'\equiv D_0 -\omega$ and we use a rotating wave approximation (RWA).
In the real experiments, we can easily change the frequency of the microwave driving while the dynamical control of the zero-field splitting is difficult. So we assume that $D_0$ is constant while we change $\omega$ during QA. 
Throughout of our paper, we set the following.
\begin{eqnarray}
&& D'=-2D'_0(t-\frac{T}{2})/T \\
&& \lambda_x(t)=B e^{-(t-\frac{T}{2})^2/\sigma ^2}
\end{eqnarray}
At $t=0$ ($t=T$), the ground state
of the Hamiltonian after the RWA is approximately described
by
$|\psi _0\rangle =\bigotimes _{j=1}^L|0\rangle _j$
 for $\sigma \gg T$.
On the other hand, at $t=T$, degenerate ground states
of the Hamiltonian after the RWA are described
by
$|{\rm{GHZ}}_{\pm}
\rangle =\frac{1}{\sqrt{2}}\bigotimes _{j=1}^L|1\rangle _j\pm \frac{1}{\sqrt{2}}\bigotimes _{j=1}^L|-1\rangle _j
$ for $\lambda_x(t)=0$
and $E_x= 0$.
We plot an energy diagram of the Hamiltonian \eqref{rwah} in Fig \ref{energydiagram}, and we confirm that the energy gap between the ground state and first excited state becomes smaller
as the time $t$ approaches to $T$.

Importantly, the Hamiltonian in the Eq. \eqref{rwah} commutes with a parity operator of $\hat{P}=\bigotimes _{j=1}^L (
|B\rangle _j\langle B|
-|D\rangle _j\langle D|
+|0\rangle _j\langle 0|
)$, and we have $\hat{P}|\psi _0\rangle=|\psi _0\rangle$ while we have $\hat{P}|{\rm{GHZ}}_{\pm} \rangle= (\pm 1 )|{\rm{GHZ}}_{\pm}\rangle $.
Therefore, by preparing a state of $|\psi _0\rangle$, the adiabatic change in the Hamiltonian allows us to create the state of $|{\rm{GHZ}}_{+} \rangle$ where non-adiabatic transitions between $|{\rm{GHZ}}_{+} \rangle$ and $|{\rm{GHZ}}_{-} \rangle$ are prohibited due to the difference of the symmetry.

\begin{figure}[h!]
  \includegraphics[width=0.45\textwidth]{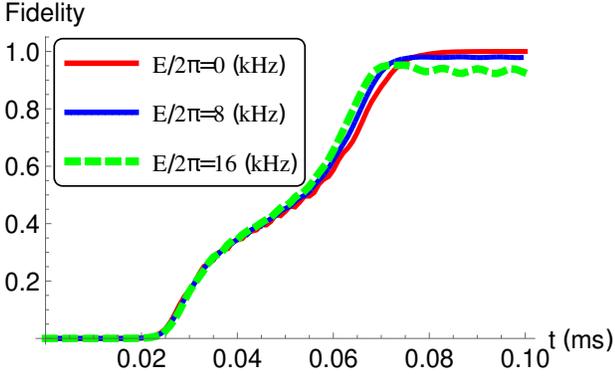}
  \caption{We plot a fidelity 
  against the time with the unit of milliseconds where we adopt the Hamiltonian without a rotating wave approximation. We set the parameters as $L=2$
  $D_0'/2\pi= 200$ kHz, $E_x^{(1)}/2\pi= E$ kHz, $E_x^{(2)}/E_x^{(1)}= 1.2$, $J_{12}/2\pi=30$ kHz, $J_{12}'/2\pi=60$ kHz, $B/2\pi=340$ kHz, $\sigma =0.2 T$, $T=0.1$ ms,  $\gamma=0$, and $\omega=40$ MHz.
  }
  \label{escalefidelityrwa}
\end{figure}

\section{Numerical simulations to generate GHZ states with bifurcation-based quantum annealing}

To evaluate the performance of our scheme, we perform numerical simulations 
to plot the fidelity between the target GHZ state and the state after QA. Here, we adopt the Hamiltonian in the Eq. \eqref{hnorwa}. 
To consider the decoherence, we use the following GKSL master equation \cite{gorini1976completely,lindblad1976generators}
\begin{eqnarray}
 \frac{d\rho }{dt}=-i[H,\rho ] +\sum_{j=1}^{L}\frac{\gamma }{2} (2\hat{L}_j \rho \hat{L}_j^{\dagger }-\hat{L}^{\dagger }_j \hat{L}_j
 \rho 
-\rho \hat{L}^{\dagger }_j \hat{L}_j
)\ 
\end{eqnarray}
where $\gamma$ denotes a decoherence rate and $\hat{L}_j$ denotes a lindblad operator at the $j$-site. Throughout of this paper, we use  $\hat{L}_j=\hat{S}_z^{(j)}$, which corresponds to magnetic field noise that is typical for the NV centers \cite{de2010universal,matsuzaki2016optically,bauch2020decoherence,hayashi2020experimental}.
We define a fidelity as $F=\langle {\rm{GHZ}}_+|\rho (t)|{\rm{GHZ}}_+ \rangle $. 

\begin{figure}[h!]
  \includegraphics[width=0.45\textwidth]{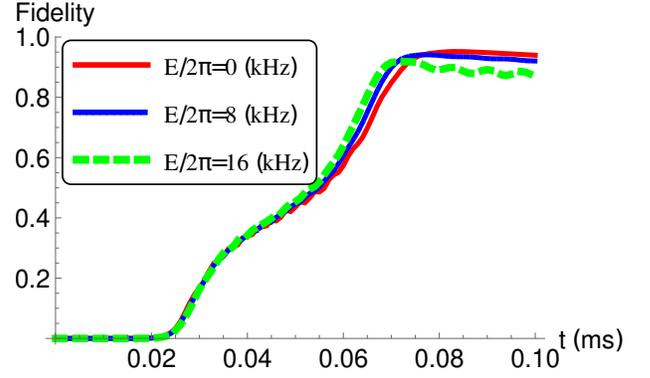}
  \caption{We plot a fidelity 
  against the time with the unit of milliseconds where we adopt the Hamiltonian without a rotating wave approximation. We set the parameters as $L=2$
  $D_0'/2\pi= 200$ kHz, $E_x^{(1)}/2\pi= E$ kHz, $E_x^{(2)}/E_x^{(1)}= 1.2$, $J_{12}/2\pi=30$ kHz, $J_{12}'/2\pi=60$ kHz, $B/2\pi=340$ kHz, $\sigma =0.2 T$, $T=0.1$ ms,  $\gamma=0.5$ kHz, and $\omega=40$ MHz.
  }
  \label{twofidedeco}
\end{figure}
We plot the fidelities against $t$ for $L=2$
without decoherence
in Fig. \ref{escalefidelityrwa}.
When there is no strain, the fidelity is more than $0.999$, and this means that the adiabatic condition is reasonably satisfied.
When we add the effect of the strain, the fidelity becomes as small as 0.979  (0.925) for $E/2\pi=8$ ($E/2\pi=16$) kHz, as shown in Fig. \ref{escalefidelityrwa}. This comes from the fact that a ground state of the Hamiltonian with the strain is not the GHZ state.
To obtain a high-fidelity GHZ state among the NV centers, it is crucial to suppress the effect of the strain by applying suitable amount of the electric fields.
In the real experiment, we have $D_0/2\pi \simeq \omega/2 \pi \simeq 2.88$ GHz. However, the computational cost becomes expensive when $D_0/2\pi$ is much larger than the other parameters.
Therefore, throughout of this paper, we set $D_0/2\pi \simeq \omega/2 \pi = 40$ MHz. Since we confirm that the dynamics does not significantly change even when we increase $D_0/2\pi$ and $\omega/2\pi$ around this parameter range, we believe that our numerical simulations are still useful to predict the experimental results for $D_0/2\pi \simeq \omega/2 \pi \simeq 2.88$ GHz.
\begin{figure}[h!]
  \includegraphics[width=0.45\textwidth]{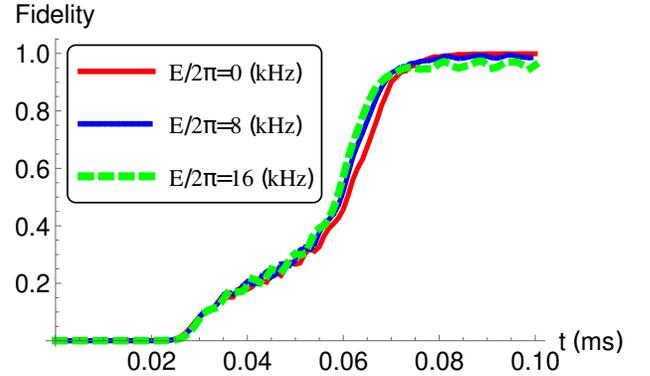}
  \caption{We plot a fidelity 
  against the time with the unit of milliseconds where we adopt the Hamiltonian without a rotating wave approximation. We set the parameters as $L=3$
  $D_0'/2\pi= 200$ kHz, $E_x^{(1)}/2\pi= E$ kHz, $E_x^{(2)}/E_x^{(1)}= 1.2$,
  $E_x^{(3)}/E_x^{(2)}= 1.2$,,
  $J_{12}/2\pi=J_{23}/2\pi=30$ kHz, $J_{12}'/2\pi=J_{23}'/2\pi=60$ kHz,
  $J_{12}/J_{13}=J'_{12}/J'_{13}=8$, $B/2\pi=340$ kHz, $\sigma =0.2 T$, $T=0.1$ ms,  $\gamma=0$, and $\omega=40$ MHz.
  }
  \label{threefide}
\end{figure}

\begin{figure}[h!]
  \includegraphics[width=0.45\textwidth]{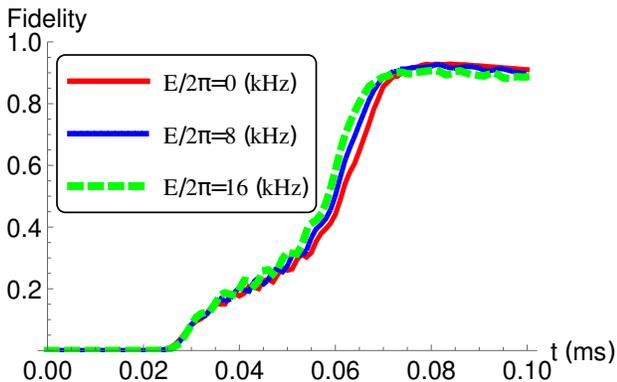}
  \caption{We plot a fidelity 
  against the time with the unit of milliseconds where we adopt the Hamiltonian without a rotating wave approximation. We set the parameters as $L=3$
  $D_0'/2\pi= 200$ kHz, $E_x^{(1)}= E$ kHz, $E_x^{(2)}/E_x^{(1)}= 1.2$,
  $E_x^{(3)}/E_x^{(2)}= 1.4$,
  $J_{12}/2\pi=J_{23}/2\pi=30$ kHz, $J_{12}'/2\pi=J_{23}'/2\pi=60$ kHz,
  $J_{12}/J_{13}=J'_{12}/J'_{13}=8$, $B/2\pi=340$ kHz, $\sigma =0.2 T$, $T=0.1$ ms,  $\gamma=0.5$ kHz, and $\omega=40$ MHz.
  }
  \label{threefiddece}
\end{figure}

Also, we plot the fidelity under the effect of decoherence 
against $t$ for $L=2$ 
in Fig. \ref{twofidedeco}.
Compared with the fidelity by using the unitary dynamics (plotted in Fig. \ref{escalefidelityrwa}), the fidely becomes smaller as expected.
However, the fidelity is still around $0.9$, and so these results show that we can generate the GHZ states even under noisy environments.

Importantly,
there was an experimental demonstration to generate an entanglement between two NV centers
\cite{dolde2013room}. However, the previous scheme requires a complicated pulse sequence, and the necessary number of the pulse operations increases as the number of NV centers increases. Moreover, the NV centers should be individually controlled by using frequency selectivity. 
On the other hand, our protocol just requires global applications of the microwave pulses without individual adressing of the NV centers, which would be beneficial to generate a GHZ states with more than two NV centers.

Finally, we plot the fidelities against $t$ for $L=3$ with and without decoherence, as shown in Fig \ref{threefide} and 
\ref{threefiddece}, respectively.
When we consider the unitary dynamics, the fidelities with $L=3$ are comparable with those with $L=2$, as shown in Fig. \ref{threefide}. This means that, for $L=3$, the adiabatic conditions are reasonably satisfied.
With decoherence, the fidelities becomes worse than those without decoherence. However, as shown in Fig. \ref{threefiddece}, the fidelities
are still around $0.9$. Again, these results show the practicality of our scheme.

\section{Conclusion}
In conclusion, we propose a scheme to generate GHZ states between spin-1 particles by using bifurcation-based QA.
Suppose that there are dipole-dipole couplings between the spin-1 particles.
After each spin-1 particle is prepared in $|0\rangle$, we slowly turn on the microwave driving, and we finally turn off the the microwave driving in an adiabatic way.
We show that
 adiabatic changes in frequency and amplitude of the microwave driving fields provide a GHZ states after QA.
 Although the energy gap between the ground state and first excited state becomes nearly degenerate when we turn off the microwave driving fields, we show that a symmerty of the Hamiltonian protects the state from the non-adiabatic transitions.
Our scheme could be useful for
possible applications to quantum information processing by using nitrogen vacancy centers in diamond.

\begin{acknowledgments}
This work was supported by MEXT's Leading Initiative for Excellent Young Researchers, KAKENHI (20H05661), and JST PRESTO (Grant No. JPMJPR1919), Japan. 
\end{acknowledgments}

~~~


\providecommand{\noopsort}[1]{}\providecommand{\singleletter}[1]{#1}%

\end{document}